\documentclass[a4paper,fleqn]{cas-sc}

\usepackage[numbers,sort&compress]{natbib}
\usepackage{graphicx}
\usepackage{amsmath,amssymb}
\usepackage{booktabs}
\usepackage{array}
\usepackage{textcomp}
\usepackage{float}
\usepackage{xcolor}

\begin{document}
\let\WriteBookmarks\relax
\def\floatpagepagefraction{1}
\def\textpagefraction{.001}

\shorttitle{The giant pulse population of PSR B0355+54}
\shortauthors{S. L. Kurdubov and D. A. Marshalov}

\title[mode=title]{The giant pulse population of PSR B0355+54}

\author[1]{S. L. Kurdubov}
\author[1]{D. A. Marshalov}

\affiliation[1]{organization={Institute of Applied Astronomy of the Russian Academy of Sciences},
  addressline={Kutuzova embankment 10},
  city={St Petersburg},
  postcode={191187},
  country={Russia}}

\begin{abstract}
Giant radio pulses are rare bright radio bursts that occur in restricted ranges
of pulsar rotational phase. Here we report giant pulses from PSR B0355+54,
a pulsar with spin period $P\simeq0.156$~s. Using 7.97 hours of observations
centred at 1.46~GHz, with 128~MHz bandwidth per circular polarization,
we identify 432 pulse periods containing bright pulses. The giant pulses
recur in two compact longitude regions inside the radio emission window.
They are narrow compared with the mean profile, with median $W_{50}=290.3\,\mu{\rm s}$,
and reach relative peak flux density ratios up to $S_{\rm pk}/\langle S_{\rm pk}\rangle=149.7$.
The early longitude group has a timing scatter of $139.7\,\mu{\rm s}$,
or $8.9\times10^{-4}$ of a rotation. The first longitude group favours right circular polarization,
and the second favours left circular polarization.
\end{abstract}

\begin{keywords}
pulsars: individual: PSR B0355+54 \sep pulsars: general \sep giant pulses \sep methods: data analysis
\end{keywords}

\maketitle

\section{Introduction}
\label{sec:introduction}
Giant radio pulses (GPs) are rare individual radio pulses that stand out from
normal pulse to pulse modulation by high intensity, short duration and
restricted rotational longitude.  A peak amplitude threshold alone gives an incomplete
definition.  The useful test is whether the
bright events form a phase confined emission state, with narrow widths and a
high intensity tail, or form the brightest part of a broader normal single
pulse distribution \cite{Cairns2004GP,Knight2006GBT,Weltevrede2006B0656RRAT}.

The cleanest GP sources are phase confined.  In the Crab pulsar, GPs occur at
specific pulse longitudes, have broad intensity distributions and can show
nanosecond time structure
\cite{StaelinReifenstein1968,Lundgren1995,Hankins2003,PopovStappers2007}.  In
PSR B1937+21, GPs occur in narrow windows tied to the regular main pulse and
interpulse, and can contain nanosecond structure
\cite{Cognard1996,KinkhabwalaThorsett2000,Soglasnov2004,McKee2019}.  The
millisecond pulsars PSR B1821--24A and PSR J0218+4232 give related examples in
which narrow GP windows are linked to high energy pulse phases
\cite{RomaniJohnston2001,Bilous2015B1821,Knight2006GBT}.  These sources set the observational template: GP activity is identified by
the way rare bright pulses select special rotational longitudes, not by
brightness alone.

The reported census of GP sources remains small and heterogeneous.  Beyond the
classical sources, GP emission or strong GP candidates have been reported from
young or millisecond pulsars such as PSR J1823--3021A and PSR B0540--69
\cite{Knight2005Search,Knight2007J1823,JohnstonRomani2003,Johnston2004B0540}.
Reported non millisecond pulsars include PSR B1112+50, PSR B0031--07, PSR
J1752+2359, PSR B0950+08 and PSR J1047--6709
\cite{ErshovKuzmin2003B1112,KuzminErshov2004,ErshovKuzmin2005,
Kuiack2020B0950,KazantsevBasalaeva2022,Sun2021J1047}.

The non millisecond part of the census is the least uniform.  Several reported
cases were detected at low radio frequencies, where propagation, profile
evolution and different intensity measures complicate comparison with the
classical GP phenomenology at high radio frequency.  Some very bright pulse sources lie outside the secure phase confined GP class: in PSR B0656+14, strong bursts were later
linked to an extended bright pulse or RRAT like distribution, and in PSR
J0529--6652 the detectable strong pulses occur near the integrated profile peak
with widths comparable to the mean profile \cite{Weltevrede2006B0656,
Weltevrede2006B0656RRAT,Crawford2013J0529}.  PSR B0950+08 and PSR B1112+50 are
also more ambiguous than the classical sources because their strong pulses are
closely tied to normal profile components or broad parts of the radio window
\cite{ErshovKuzmin2003B1112,Kuiack2020B0950}.  Among previously reported non
millisecond GP sources, PSR J1047--6709 is the main L band comparison
\cite{Sun2021J1047}.

The data analysed here were obtained in a broader programme of high time
resolution single pulse observations with the Quasar VLBI network
\cite{KurdubovMarshalov2022JAI}.  The present paper concerns PSR B0355+54, a
pulsar with spin period \(P\simeq0.156\)~s
\cite{TaylorManchesterLyne1993,ManchesterEtAl2005}.  Its radio emission is
broad, structured and variable.  Scintillation measurements show time dependent
propagation effects, and recent L band observations show profile variability
between epochs and across frequency
\cite{Xu2018B0355Scint,Jiang2024B0355Profile}.

PSR B0355+54 is also detected in X rays.  \textit{XMM-Newton} and
\textit{Chandra} observations reveal emission from the pulsar and extended
emission interpreted as a pulsar wind nebula or bow shock structure
\cite{McGowan2006B0355PWN,Tepedelenlioglu2007B0355PWN}. 

Here we report that PSR B0355+54 emits phase confined GPs at L band.  The
selected bright pulses recur in two compact longitude regions within the radio
emission window.  They form discrete phase concentrations and do not trace the
radio window as a smooth extension of the normal profile.

\section{Observations and data reduction}
\label{sec:analysis}

\subsection{Observations and instrumental setup}
\label{sec:obs_setup}

PSR B0355+54 was observed with the Badary 32~m radio telescope of the Quasar
VLBI network~\cite{Shuygina2019} during the RUP112 and RUP117 sessions on 2026 March
25 and April 30. The observations were centred at 1464.49~MHz. A VLBI backend
recorded 2 bit raw voltage data in sixteen 16~MHz channels, split equally
between RCP and LCP, giving 128 MHz of bandwidth per polarization. A hydrogen
maser provided the station time and frequency reference. The primary sample
contains 17 scans and 28676~s of data on source. The full instrumental setup is
given in Table~\ref{tab:obssetup}.

\begin{table}
\centering
\caption{Instrumental setup for the PSR B0355+54 observations.}
\label{tab:obssetup}

\begin{tabular}{ll}
\hline
Parameter & Value \\
\hline
Observing sessions & RUP112, RUP117 \\
Station & Badary 32~m radio telescope \\
Network & Quasar VLBI network \\
Source & PSR B0355+54 / J0358+5413 \\
Frequency band & L band \\
Central frequency & \(1464.49\) MHz \\
Backend & VLBI backend \\
Recording format & Mark 5B \\
Quantization & 2 bit sampling \\
Channels per polarization & \(8\times16\) MHz \\
Recorded bandwidth & \(128\) MHz per polarization \\
Polarizations & RCP and LCP \\
Time/frequency reference & Hydrogen maser \\
Number of used scans  & 17 \\
Cumulative observation time & \(28676\) s = \(7.97\) h \\
\hline
\end{tabular}

\end{table}

\subsection{Folding and single pulse measurements}
\label{sec:pipeline}

The first bright pulse candidates were found with the coherent single pulse
search pipeline developed for the Quasar VLBI stations
\cite{KurdubovMarshalov2022JAI}.  The final population analysis used folded
single rotation profiles produced with \textsc{dspsr}
\cite{vanStratenBailes2011DSPSR}.  The strongest discovery stage event was also
reprocessed over a grid of trial dispersion measures and time resolutions; this
check is described in Appendix~\ref{app:auxiliary}.

We folded the Mark 5B recordings with coherent dedispersion at the catalogue
dispersion measure of PSR B0355+54.  The folding used 2048 phase bins per pulsar
period and 256 frequency channels per recorded 16~MHz channel.  The main
\textsc{dspsr} options were \texttt{-d 2 -s -A -b 2048 -F 256 -K}.  Each archive
subintegration corresponds to one stellar rotation and contains detected power
profiles for the two recorded circular polarization streams.  For the folding
period used in the reduction, \(P\simeq0.1563987\,\mathrm{s}\), the phase bin
duration is
\begin{equation}
\Delta t = \frac{P}{2048}=76.37\,\mu\mathrm{s}.
\end{equation}

The folded products were written as PSRCHIVE archives
\cite{vanStraten2012PSRCHIVE}.  Persistent radio frequency interference and
instrumental artefacts were inspected with \textsc{psrzap}.  The same fixed mask
for contaminated channels was then applied to every scan and polarization in the
Python analysis.

The Python pipeline read the folded archives scan by scan, applied the fixed
channel mask, kept individual stellar rotations as separate subintegrations,
estimated the off pulse baseline and noise level, and measured candidate pulse
properties.  For each candidate it records the scan identifier, polarization,
subintegration number, peak phase bin, peak phase, folded profile peak SNR,
pulse width and profile normalized peak flux density ratio.  The two circular
polarizations are measured independently, and coincidences are identified only
after the single polarization event table has been constructed.

For each scan and polarization, the pipeline also constructs a scan local clean
mean profile.  Rotations containing candidate bright pulses and rotations marked
as RFI contaminated are excluded from the average.  The remaining rotations
define the clean reference profile used for local normalization.

\subsection{Pulse strength measurement and approximate flux scale}
\label{sec:fluxscale}

Candidate strength is reported primarily through a dimensionless, scan local relative peak
flux density,
\begin{equation}
\frac{S_{\rm pk}}{\langle S_{\rm pk}\rangle}=\frac{P_{{\rm pk,pulse}}-P_{{\rm base,pulse}}}{
P_{{\rm pk,clean}}-P_{{\rm base,clean}}},
\end{equation}
where the denominator is measured from the clean mean profile for the same scan and
circular polarization.  This normalization follows local changes in sensitivity and
polarization dependent profile shape.  It is a peak flux density
ratio, not a pulse energy ratio.

We estimated peak flux densities using single polarization SEFD values of 365 Jy for RCP and 340 Jy for LCP obtained from station measurement.  
The strongest event by relative peak flux
density reaches a summed two polarization peak scale of $\simeq96$ Jy, and
the maximum SNR event reaches $\simeq120$ Jy.  The corresponding 
approximate fluence scale depends on pulse shape. The derivation and the B1937+21 SEFD consistency check are
given in Appendix~\ref{app:flux} and Table~\ref{tab:sefd}.

\subsection{Local timing refinement}
\label{sec:timing_methods}

The first folding pass used the catalogue ephemeris for PSR J0358+5413 /
B0355+54.  The astrometric and spin parameters were taken from the ATNF Pulsar
Catalogue \cite{ManchesterEtAl2005} entry based on the timing solution of Li et
al. \cite{lwy+16}.  This ephemeris is adequate within individual scans, but the
data span two observing sessions separated by several tens of days.  A small
spin frequency error can produce a measurable drift of bright pulse
longitude between scans.

We constructed a local timing refinement from the early bright pulse group.  A
local template was formed from GP profiles aligned to their measured peaks.  For
each early group GP, a time of arrival was measured by cross correlating the
single rotation profile with this template over a fixed window.  The resulting
TOAs were fitted with \textsc{tempo2}
\cite{HobbsEtAl2006,EdwardsEtAl2006}.

Only the spin frequency $F_0$ was adjusted in the timing model.  The local fit
used 139 TOAs from the early group.  The prefit residual RMS was
$8514.5\,\mu\mathrm{s}$, and the postfit residual RMS was
$140.9\,\mu\mathrm{s}$.  The fitted spin frequency changed from
$F_0=6.39440176623564\,\mathrm{s}^{-1}$ to
$F_0=6.39440180469109\,\mathrm{s}^{-1}$, with a formal uncertainty of
$5.38\times10^{-11}\,\mathrm{s}^{-1}$ at PEPOCH = 61142.  The refined spin
frequency was then used to refold the data and define the corrected pulse
longitude scale.  The timing scatter quoted in Table~\ref{tab:key_measurements}
was computed after applying the same residual clipping used in the local fit.

\section{Full period giant pulse search}
\label{sec:full_period_search}

\subsection{Blind event selection and background estimate}
\label{sec:event_selection}

The candidate search was deliberately blind in pulse longitude.  After
application of the fixed frequency channel mask, each single rotation profile
was searched across all 2048 phase bins in each circular polarization.  A
candidate entered the clean event table if it passed the RFI veto and its folded
profile peak signal to noise ratio satisfied \({\rm SNR}\ge5.4\).  No cut on
pulse width or relative peak flux density was applied.  The selected events were
then grouped by pulse phase for morphological analysis.  The early and late
groups are not search gates; they describe the two repeatable longitude
concentrations that appear after the full period selection.

Because the folded profiles are detected power profiles, the off pulse statistic
is better described as an averaged power statistic than as a voltage amplitude
statistic.  For an effective bandwidth \(\Delta\nu_{\rm
eff}=128\,\mathrm{MHz}\times204/256\simeq102\) MHz and a phase bin duration
\(\Delta t=76.37\,\mu\mathrm{s}\), the number of independent power samples in
one profile bin is approximately
\begin{equation}
n\simeq\Delta\nu_{\rm eff}\Delta t\simeq7.8\times10^3.
\end{equation}
The corresponding detected power statistic can be approximated as
\begin{equation}
X\sim \frac{\chi^2_{2n}}{2n}.
\end{equation}
For a threshold of \(\mathrm{SNR}=5.4\), measured relative to the off pulse
rms, the ideal single bin upper tail probability is
\begin{equation}
p_{\chi^2}=P\left[\chi^2_{2n}>2n+5.4\sqrt{4n}\right]\simeq5.9\times10^{-8}.
\end{equation}
A typical 1783 s scan contains about \(1.14\times10^4\) stellar rotations, or
\(2.3\times10^7\) phase bin trials per polarization, giving an ideal expectation
of \(N_{\rm out,scan,pol}\simeq1.4\) detected power outliers per scan per
polarization.

Across the primary sample, using \(1.83\times10^5\) rotations, two circular
polarizations and about 2000 phase bins outside the two longitude groups, the
same ideal calculation gives \(\simeq43\) single bin outliers outside the two
groups.  The observed background sample outside the two longitude groups
contains 63 candidates, comparable to this expectation once channel masking,
correlated bins, baseline estimation, residual RFI and leakage from normal
emission are allowed for.  This background is small compared with the 551 events
in the two longitude groups, and the background sample outside the groups
contains no coincidences between RCP and LCP.

\subsection{Crab pulsar reference processed with the same selection}
\label{sec:crab_reference}

The Crab pulsar was used as a reference processed with the same automatic
selection.  The selected Crab events recover the expected dominant main pulse
concentration, wrapping through phase zero, together with a smaller interpulse
concentration.  Of the 162 clean Crab events, 142 are in the main pulse region,
9 in the interpulse region and 11 outside these two regions.

The Crab sample is more extreme than PSR B0355+54 in recurrence rate,
polarization coincidence fraction, peak flux density ratio and pulse width
(Table~\ref{tab:crab}).  This is the expected behaviour for a classical GP source.  The Crab scan
provides a reference for a known, more extreme GP population processed with the
same selection.

\begin{table}[H]
\centering
\caption{Comparison between the PSR B0355+54 primary sample and the Crab reference
processed with the same selection.}
\label{tab:crab}
\footnotesize
\begin{tabular}{lrr}
\hline
Quantity & PSR B0355+54 & Crab  \\
\hline
Analysed interval (s) & 28676 & 630.2 \\
Analysed rotations & \(1.83\times10^5\) & \(1.887\times10^4\) \\
Selected events & 551 & 162 \\
Pulse periods with GP & 432 & 101 \\
Rotations / GP & 424 & 187 \\
Median \(S_{\rm pk}/\langle S_{\rm pk}\rangle\) & 63.6 & 251.9 \\
Maximum \(S_{\rm pk}/\langle S_{\rm pk}\rangle\) & 149.7 & 1170.4 \\
Median \(W_{50}\) (\(\mu{\rm s}\)) & 290.3 & 24.9 \\
\hline
\end{tabular}
\end{table}

\section{Results}
\label{sec:results}
\subsection{Phase confinement}
\label{sec:phase_confinement_results}

A search over the full pulse period identifies repeatable giant pulses in
PSR B0355+54. Across the 7.97~h primary data set, 432 stellar
rotations contain selected bright pulses, giving one GP rotation per
\(\simeq424\) rotations.  Counting the right and left circular polarizations (RCP and LCP)
separately, the primary sample contains 551 events.  The events appear in both
observing sessions and in both circular polarizations.  The event selection and background estimate are given in Section~\ref{sec:event_selection}.
Selected events are present in all used scans; the scan level distribution is
listed in Table~\ref{tab:scans}.

\begin{figure}
\centering
\includegraphics[width=\linewidth]{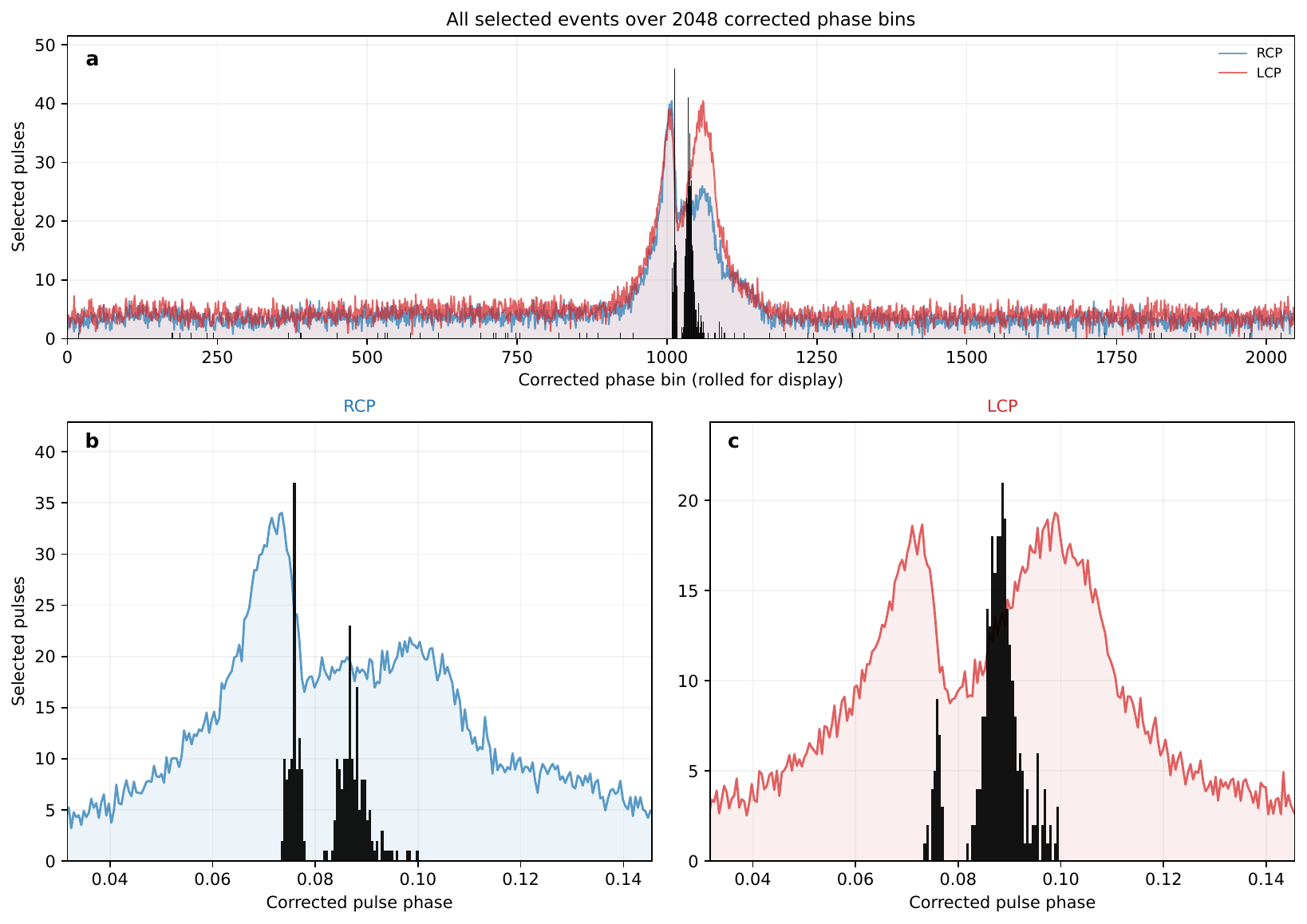}
\caption{Phase confinement of the selected pulses.
\textbf{a}, Full 2048 bin corrected pulse phase histogram,
with the mean profiles overlaid for RCP and LCP.  The candidates
outside the two main longitude groups are sparse and spread over phase.
\textbf{b,c}, Zooms of the main pulse region for RCP and LCP.  The selected
pulses cluster into two preferred longitude groups within the radio emission
window.}
\label{fig:phase_confinement}
\end{figure}

\begin{table}
\centering
\caption{Key measurements of the PSR B0355+54 GP population.
Events are counted separately in RCP and LCP.  GPs are
counted once for each pulsar period and longitude group.  A GP is counted as
detected in both polarizations when selected RCP and LCP events occur in the
same pulsar period and the same longitude group.  Timing scatter is measured
from the post fit residuals after outlier clipping.}
\label{tab:key_measurements}
\begin{tabular}{lccc}
\toprule
Measurement & Full sample & Early group & Late group \\
\midrule
Polarization resolved events & 551 & 139 & 412 \\
GPs & 432 & 113 & 319 \\
RCP / LCP events & 265 / 286 & 108 / 31 & 157 / 255 \\
RCP only GPs & 146 & 82 & 64 \\
LCP only GPs & 167 & 5 & 162 \\
RCP\&LCP GPs & 119 & 26 & 93 \\
Median pulse phase & -- & 0.07593 & 0.08813 \\
Median \(W_{50}\) (\(\mu\)s) & 290.3 & 320.5 & 271.4 \\
Median \(S_{\rm pk}/\langle S_{\rm pk}\rangle\) & 63.6 & 57.5 & 65.5 \\
Maximum \(S_{\rm pk}/\langle S_{\rm pk}\rangle\) & 149.7 & 127.9 & 149.7 \\
Timing scatter \(\sigma_t\) (\(\mu\)s) & -- & 139.7 & 474.5 \\
\bottomrule
\end{tabular}
\end{table}

\begin{table*}
\centering
\caption{Scan level summary of the uniform automatic PSR B0355+54 bright pulse
sample.  Session distinguishes the two observing blocks, since scan numbers
are not unique across sessions.  RCP and LCP are event counts resolved by polarization.
Unique rotations count pulse periods containing at least one selected event.}
\label{tab:scans}

\begin{tabular}{llrrrrrrr}
\hline
Session & Scan & RCP & LCP & Uniq. rot. & Both pol & Scan len. & Max & Median \\
        &      &     &     &            &          &  (s) &  $S_{\rm pk}/\langle S_{\rm pk}\rangle$ & $W_{50}$ ($\mu$s) \\
\hline
rup112 & no0001 & 10 & 16 & 17 & 9 & 1783 & 107.0 & 307.7 \\
rup112 & no0002 & 4 & 7 & 9 & 2 & 1784 & 94.6 & 336.8 \\
rup112 & no0003 & 2 & 8 & 10 & 0 & 1649 & 82.5 & 195.5 \\
rup112 & no0036 & 27 & 25 & 44 & 8 & 1783 & 94.2 & 283.0 \\
rup112 & no0037 & 30 & 39 & 52 & 17 & 1784 & 134.4 & 258.4 \\
rup112 & no0038 & 17 & 15 & 24 & 8 & 1549 & 127.9 & 284.6 \\
rup112 & no0040 & 25 & 19 & 35 & 9 & 1784 & 108.3 & 303.6 \\
rup112 & no0041 & 18 & 20 & 29 & 9 & 1784 & 88.2 & 331.9 \\
rup112 & no0042 & 15 & 17 & 26 & 6 & 1784 & 99.7 & 232.3 \\
rup112 & no0043 & 21 & 22 & 36 & 7 & 1784 & 112.9 & 309.7 \\
rup117 & no0001 & 27 & 22 & 35 & 14 & 1738 & 126.2 & 281.8 \\
rup117 & no0002 & 23 & 22 & 36 & 9 & 1738 & 147.0 & 303.2 \\
rup117 & no0003 & 13 & 21 & 27 & 7 & 1739 & 140.8 & 289.1 \\
rup117 & no0004 & 12 & 6 & 14 & 4 & 1738 & 136.8 & 321.6 \\
rup117 & no0005 & 12 & 17 & 22 & 7 & 1739 & 149.7 & 303.8 \\
rup117 & no0006 & 5 & 4 & 9 & 0 & 958 & 73.5 & 312.6 \\
rup117 & no0010 & 4 & 6 & 7 & 3 & 1559 & 111.1 & 227.3 \\
All & Total & 265 & 286 & 432 & 119 & 28676 & 149.7 & 290.3 \\
\hline
\end{tabular}

\end{table*}

The selected pulses are distinguished first by their longitude distribution.
They concentrate into two longitude groups superposed on the normal profile
(Fig.~\ref{fig:phase_confinement}).  The two groups are separated by 1.91 ms in
pulse longitude.  The late group contains most of the events, and the early
group is rarer.

\subsection{Circular polarization structure}
\label{sec:polarization_results}

The polarization information adds structure to this longitude picture.  Of the
432 bright pulse rotations, 119 contain selected events in both circular
polarizations, and 313 are selected in only one.  These RCP only and LCP only
detections are a large part of the sample and carry information about the
polarization pattern.  Their occurrence depends strongly on longitude.  In the
early group, 82 rotations are RCP only, 5 are LCP only, and 26 are detected in
both circular polarizations.  In the late group, 64 rotations are RCP only,
162 are LCP only, and 93 are detected in both circular polarizations.  When both
polarizations are selected in the same rotation, the two events always fall in
the same longitude group (Table~\ref{tab:key_measurements}).

\subsection{Pulse width and relative strength}
\label{sec:width_strength_results}

\begin{figure}
\centering
\includegraphics[width=\linewidth]{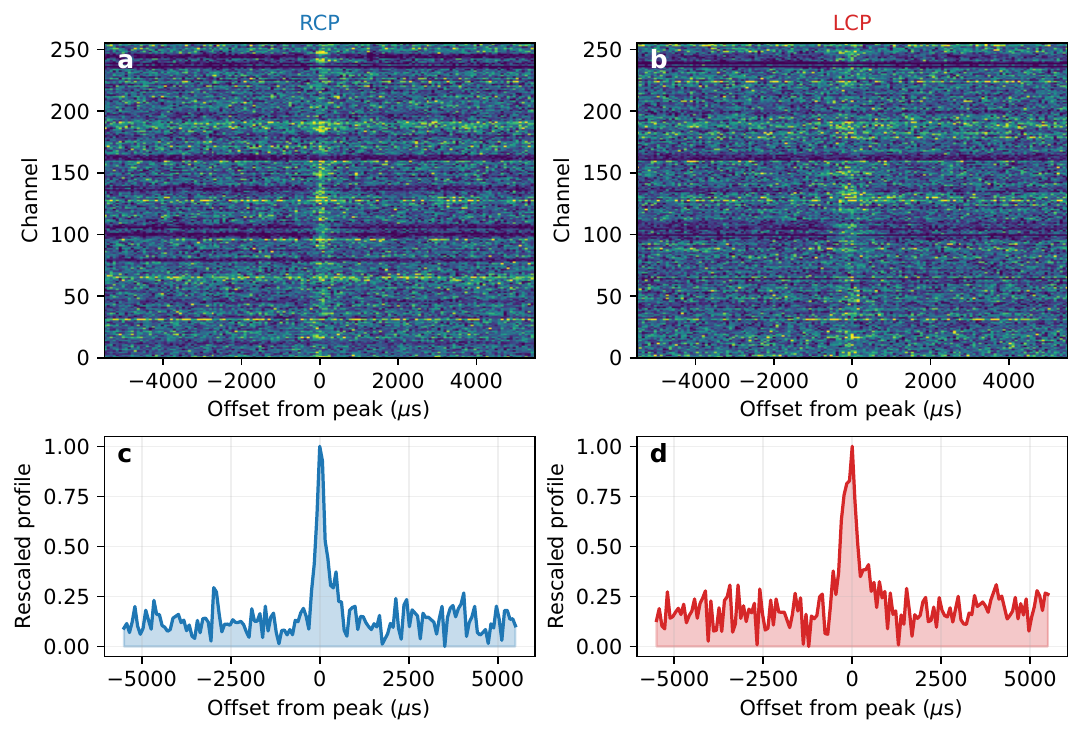}
\caption{Representative narrow bright pulse.  \textbf{a,b}, Dynamic spectrum
cutouts around the representative saved event in RCP and LCP, shown before
application of the fixed channel mask.  \textbf{c,d}, Corresponding folded event
profiles aligned to their own peaks.  The pulse is broadband, narrow on the
folded profile scale and detected in both circular polarizations.}
\label{fig:representative_event}
\end{figure}

The selected pulses are also narrow compared with the normal emission.  Their
median width is \(W_{50}=290.3\,\mu{\rm s}\), much narrower than the mean radio profile, 
and their relative peak flux densities extend to \(S_{\rm pk}/\langle S_{\rm pk}\rangle=149.7\).  
Figure~\ref{fig:representative_event} shows a representative narrow event, 
rup117 scan no0001, subintegration 7727, detected in both circular polarizations with broadband dynamic spectrum
structure.

\begin{figure}
\centering
\includegraphics[width=0.82\linewidth]{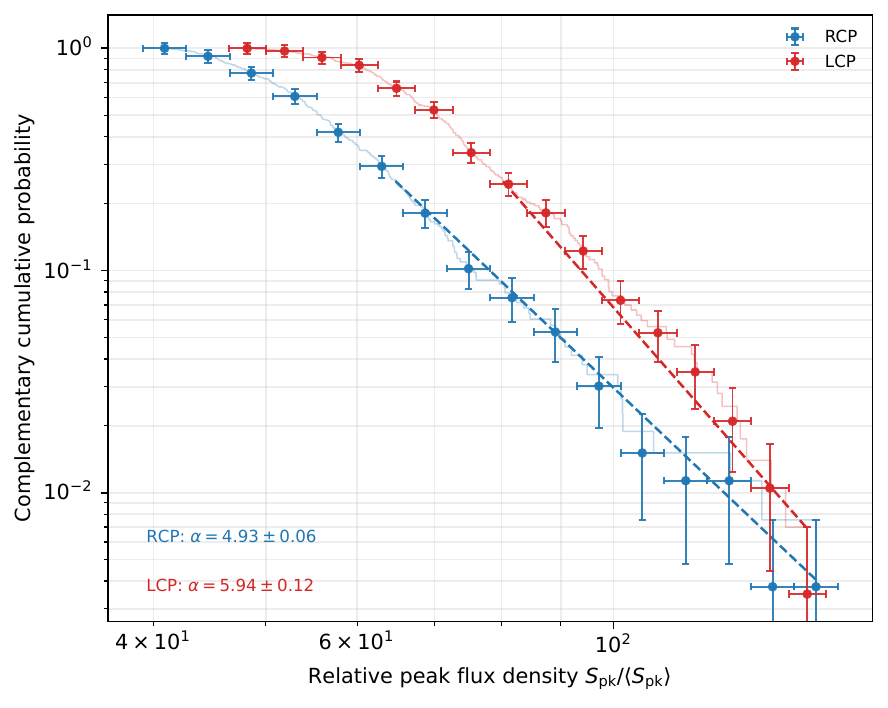}
\caption{Relative peak flux density distribution of the GP sample.
Points show empirical complementary cumulative probabilities evaluated at
logarithmically spaced thresholds. Error bars show counting uncertainties,
including horizontal threshold intervals.  Dashed lines show formal descriptive
fits to the terminal high intensity segments.}
\label{fig:ccdf}
\end{figure}

Both circular polarizations show extended high intensity tails
(Fig.~\ref{fig:ccdf}).  Formal descriptive fits to the upper quartile of the
complementary cumulative distributions quantify the terminal tails with slopes
of \(4.93\pm0.06\) for RCP and \(5.94\pm0.12\) for LCP.

\begin{figure}
\centering
\includegraphics[width=\linewidth]{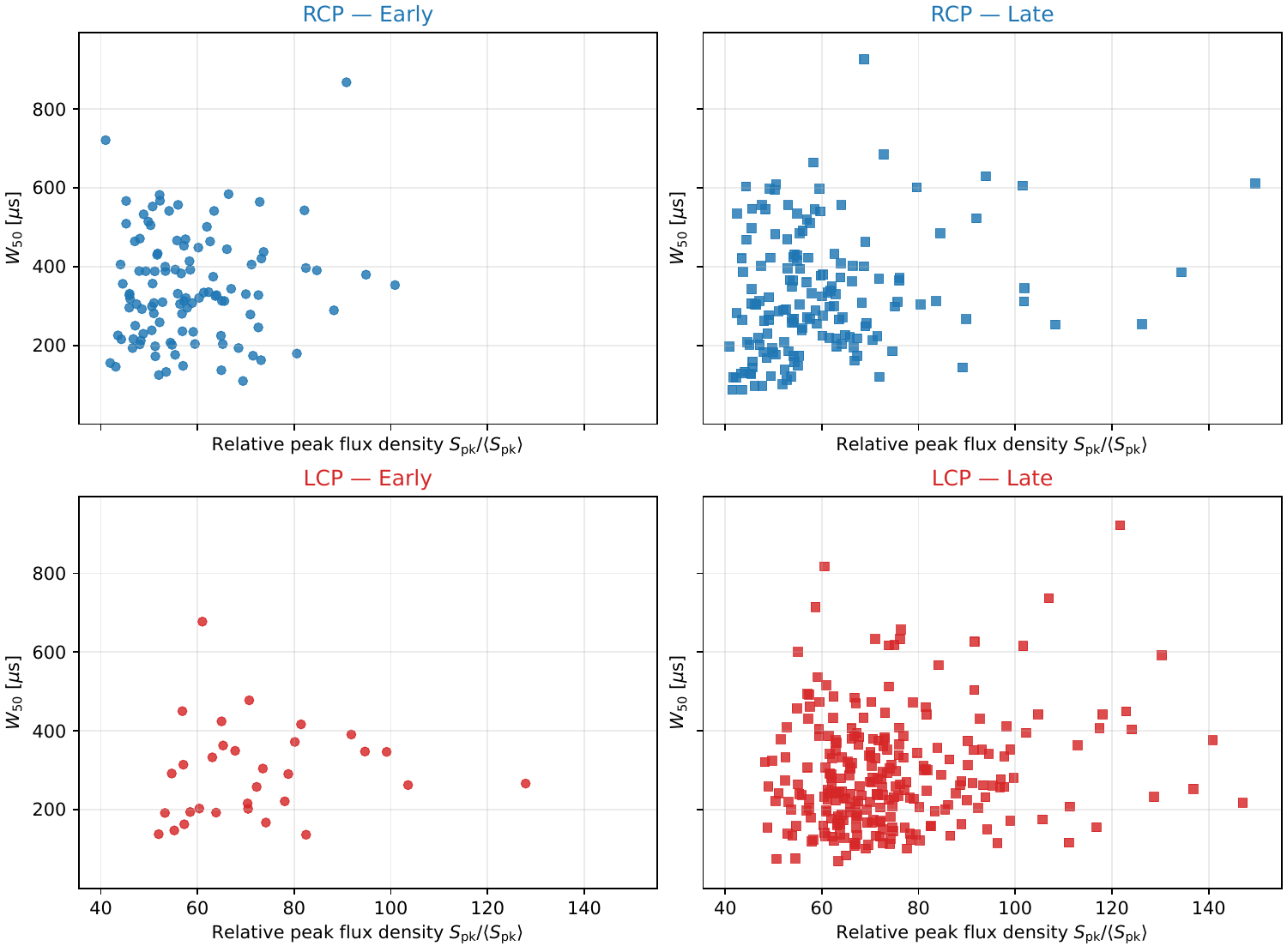}
\caption{Pulse width versus relative peak flux density, shown separately for each circular polarization and longitude group.}
\label{fig:width_strength}
\end{figure}

Figure~\ref{fig:width_strength} shows the same width measurement as a function of
relative peak flux density for each circular polarization and longitude group.
The selected pulses occupy a narrow range of folded profile widths over the full
observed range of relative peak strength.

\subsection{Timing compactness}
\label{sec:timing_results}

\begin{figure}
\centering
\includegraphics[width=\linewidth]{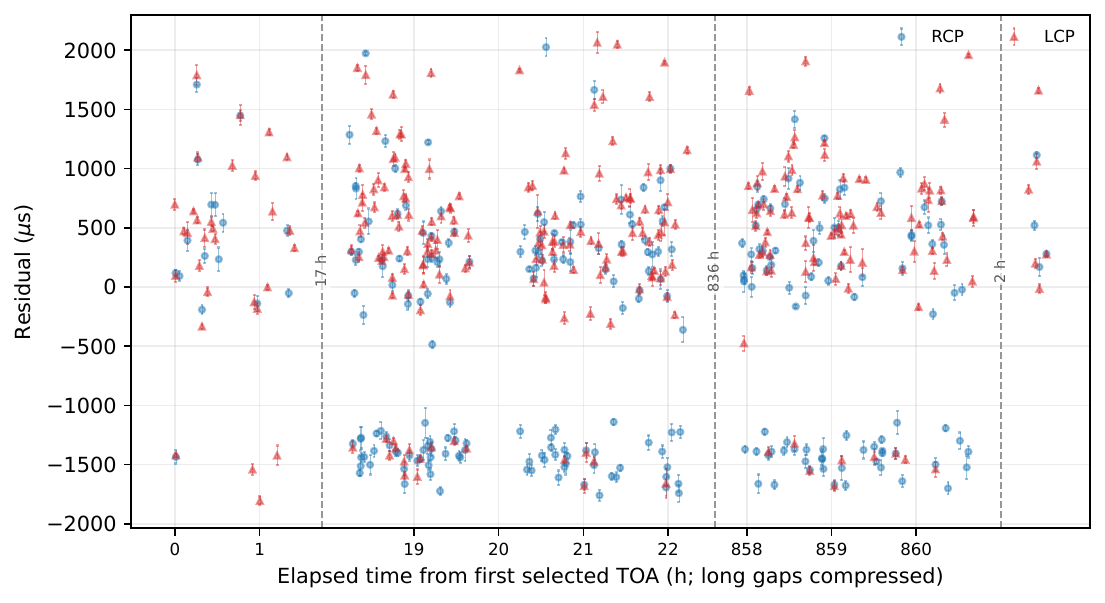}
\caption{Timing compactness of the selected bright pulse groups.  Post fit timing residuals are shown versus observing time, with long gaps compressed and labelled.  Points show individual selected events with their timing uncertainties. Colours and markers distinguish the two circular polarizations.  The two residual clouds correspond to the early and late phase groups, with the early group confined to about $10^{-3}$ of the stellar rotation.}
\label{fig:timing_compactness}
\end{figure}

The early longitude group is the most compact in phase.  Times of
arrival formed from the selected pulses give a post fit scatter of
\(139.7\,\mu\mathrm{s}\) for the early group, about \(10^{-3}\) of a rotation,
compared with \(474.5\,\mu\mathrm{s}\) for the broader late group
(Fig.~\ref{fig:timing_compactness}, Table~\ref{tab:key_measurements}).  Together, the
selected events show three linked signatures of a phase confined population:
they recur in two compact longitude groups, 
detections in RCP and LCP remain within the same group when both are present, and the
control region outside the groups contains only sparse background candidates
with no repeated phase bin and no comparable coincidence between RCP and LCP.

\section{Discussion}
\label{sec:discussion}
\subsection{Comparison with classical giant pulse sources}
\label{sec:discussion_comparison}

PSR B0355+54 falls on the phase confined side of the GP phenomenology.
The selected bright pulses do not form a smooth high amplitude extension across
the radio window.  They recur at two preferred longitudes inside the normal
pulse window, and the full period search leaves only a sparse background outside
those groups.  This differs from bright pulse or RRAT like cases in which the strongest
pulses mainly trace the normal emission window
\cite{Weltevrede2006B0656RRAT,Crawford2013J0529}.

The closest comparison is PSR B1937+21.  The analogy is geometric and is best made in rotational phase.
In PSR B1937+21, giant pulses occur in narrow phase regions
associated with the trailing edges of the regular main pulse and interpulse
components \cite{KinkhabwalaThorsett2000,McKee2019}. 
For B1937+21, the mean widths of the folded main pulse and interpulse GP
distributions are about \(4\,\mu{\rm s}\), corresponding to
\(\simeq 2.6\times10^{-3}\) of the rotation \cite{McKee2019}.  In PSR B0355+54, the early group
has a timing scatter of \(139.7\,\mu{\rm s}\), or
\(8.9\times10^{-4}\) of the period, and the broader late group has a scatter
of \(3.0\times10^{-3}\) of the period.  In both objects, rare high peak pulses select longitudes tied to the mean radio profile.  
In PSR B0355+54 the early group lies near the trailing side of the first profile component, and the broader
late group is associated with the second component.

The polarization behaviour gives a second parallel with PSR B1937+21.  Strong
giant pulses from PSR B1937+21 have been detected above threshold in either RCP
or LCP without simultaneous threshold crossing in both 
channels \cite{PopovSoglasnov2004B1937Polarization}.  Later full polarization
studies showed that individual B1937+21 giant pulses can have high linear or
circular polarization, unlike the average profile
\cite{Zhuravlev2013B1937Polarization,McKee2019}.  
In PSR B0355+54, the circular polarization preference changes with longitude:
RCP only detections dominate the early group, and LCP only detections
dominate the late group.

The Crab pulsar gives a useful comparison in absolute arrival phase.  In an
L band Crab GP sample, Majid et al. measured TOA residual widths of about
\(90\,\mu{\rm s}\) for main pulse GPs and \(140\,\mu{\rm s}\) for interpulse
GPs \cite{Majid2011}.  In PSR B0355+54, the early longitude group has a comparable
timing scatter of \(139.7\,\mu{\rm s}\), and the broader late group
has \(474.5\,\mu{\rm s}\).  

Table~\ref{tab:literature} gives numerical context for reported non millisecond
GP samples.  It should not be read as a uniform taxonomy: the listed studies use
different observing frequencies, sensitivity limits, and intensity or energy
measures.  The maximum scale is only approximate.  In this comparison,
PSR B0355+54 is a rare L band case.  Its recurrence rate and peak scale are
comparable to other reported non millisecond GP sources, and its strongest
classification evidence is the full period phase confinement.

\begin{table}
\centering
\caption{Comparison with known non millisecond pulsars GP samples. The maximum scale follows
the relative intensity or energy measure reported in each reference and is
intended as an approximate comparison.}
\label{tab:literature}
\begin{tabular}{lrrrrrrl}
\toprule
Pulsar & Freq. (MHz) & Bandwidth (MHz) & Obs. time (h) & $N_{\rm GP}$ & $N_{\rm rot}/N_{\rm GP}$ & Max. scale & Reference \\
\midrule
PSR B1112+50 & 111 & 2.56 & 4.6 & 126 & $\sim150$ & $\sim80$ & \cite{ErshovKuzmin2003B1112} \\
PSR B0031--07 & 111 & 2.56 & 1.5 & 440 & $\sim250$ & $\simeq120$ & \cite{KuzminErshov2004} \\
PSR B0031--07 & 40 & 0.16 & 6.0 & 337 & $\sim800$ & $\sim400$ & \cite{KuzminErshov2004} \\
PSR J1047--6709 & 1369 & 256.0 & 0.42 & 75 & $\sim102$ & $\sim110$ & \cite{Sun2021J1047} \\
PSR B0355+54 & 1464 & 128.0 & 7.97 & 432 & $\sim424$ & $\sim149.7$ & this work \\
PSR J1752+2359 & 111 & 2.56 & 5.7 & 190 & $\sim270$ & $\sim260$ & \cite{ErshovKuzmin2005} \\
\bottomrule
\end{tabular}

\end{table}

The comparison with PSR B1937+21 also emphasizes rotation count.  Because
PSR B0355+54 has a period about two orders of magnitude longer, an equivalent
rotational sample would require much longer observations.  The present data
show recurrent GPs confined in phase.  They do not yet probe the rarest part
of the amplitude distribution.

\subsection{Relation to previous studies of PSR B0355+54}
\label{sec:discussion_previous}

Previous radio studies already suggested that PSR B0355+54 is not well
described by a stationary two component beam.  Single pulse polarimetry found
rare flaring of weak profile regions and used such behaviour to argue against a
simple inactive cone sector picture \cite{MitraSaralaRankin2008}.  High
frequency polarimetry identified B0355+54 as the prototype of a class in which
one profile component remains highly polarized and has a flatter spectrum than
the rest of the profile \cite{vonHoensbroechKijakKrawczyk1998}.  Recent L band
FAST observations show profile changes between epochs and across frequency,
consistent with a nonuniform and time dependent distribution of emitting
regions \cite{Jiang2024B0355Profile}.  The bright pulses reported here give a
direct view of the high intensity end of this nonuniform emission pattern.

\subsection{Circular polarization structure and interpretation}
\label{sec:discussion_polarization}

The two longitude groups have different rates, phase widths and circular
polarization preferences.  The early group is less frequent and more compact in phase.  The late group is
more populated and spans a wider longitude range.
The RCP to LCP ratio changes from \(82/5\) in the early group to \(64/162\) in the late group.
A single gain imbalance, leakage term or observing state would not naturally
produce such a reversal between two nearby profile components.  The contrast
points to different circular mode content, or different propagation
conditions, in the two bright pulse regions.

A simple interpretation is that the line of sight crosses two active regions
within the open field line beam.  The early region may correspond to a narrower
range of field lines or emission heights, giving a stable pulse longitude and a
preferred circular mode.  The late region may sample a broader range of
longitude, height or mode conversion in the magnetospheric plasma.  In this
picture the bright pulses are rare coherent or amplified states of selected
parts of the radio beam, not a separate beam component.

\section{Conclusions}
\label{sec:conclusions}

We have reported repeatable giant pulses from PSR B0355+54 at L band.  The
full period search identifies 432 pulse periods containing selected bright
pulses, concentrated in two compact longitude groups inside the radio emission
window.  The events are narrow compared with the mean radio profile and reach
relative peak flux density ratios up to
\(S_{\rm pk}/\langle S_{\rm pk}\rangle=149.7\).  The early longitude group has a
timing scatter of \(139.7\,\mu{\rm s}\), or \(8.9\times10^{-4}\) of a rotation.
The early and late longitude groups also show different circular polarization
preferences, with RCP only detections dominant in the early group and LCP only
detections dominant in the late group.  These properties make PSR B0355+54 a
useful non millisecond L band source for studying rare coherent emission from
selected pulse longitudes, and distinguish it from bright pulse cases in which
the strongest events mainly follow the normal radio window.

\section*{Acknowledgements}
We thank Alexey Melnikov, Andrey Mikhailov, and the technical staff of the Badary
observatory for assistance with the pulsar observations.

\section*{CRediT authorship contribution statement}
S.L.K. led the data reduction, single pulse search, statistical analysis, figure
preparation and manuscript drafting. D.A.M. contributed to the observing programme,
processing workflow, interpretation of the results and manuscript revision. Both authors
reviewed and approved the final manuscript.

\section*{Declaration of competing interest}
The authors declare that they have no known competing financial interests or personal relationships that could have appeared to influence the work reported in this paper.

\section*{Data availability}
The derived event tables, scan level summaries, clean mean profiles and figure
input data are provided as Source Data files. The raw Mark 5B voltage recordings and folded PSRCHIVE 
archives are large data products and are available from the corresponding author upon 
reasonable request, subject to storage and transfer constraints.

\section*{Code availability}
The main reduction used the open source packages
\textsc{dspsr} commit 7241e80b and \textsc{psrchive} version
2026-02-26 (commit e7a724da2), together with \textsc{tempo2} version 2022.05.1 for the
timing fit. A local patch to \textsc{dspsr} was used to reorder Mark~5B frequency channels
before the standard coherent dedispersion and folding stages.
The article plotting scripts, server search wrapper and local \textsc{dspsr} patch
are available from the corresponding author upon reasonable request.

\section*{Declaration of generative AI and AI assisted technologies in the manuscript preparation process}
During the preparation of this work the authors used ChatGPT for language editing and formatting assistance. After using this tool, the authors reviewed and edited the content as needed and take full responsibility for the content of the article.

\appendix
\section*{Appendices}
\setcounter{figure}{0}
\setcounter{table}{0}
\numberwithin{figure}{section}
\numberwithin{table}{section}
\numberwithin{equation}{section}
\section{Flux density and fluence scale}
\label{app:flux}

Pulse strength is reported primarily as
\(S_{\rm pk}/\langle S_{\rm pk}\rangle\), measured relative to the clean mean
profile for the same scan and circular polarization.  To estimate approximate
peak flux densities for the strongest events, we used single polarization SEFD
values measured at the Badary station:
\({\rm SEFD}_{\rm RCP}=365\)~Jy and \({\rm SEFD}_{\rm LCP}=340\)~Jy.

With $\Delta\nu_{\rm eff}\simeq102$ MHz and $\Delta t_{\rm
bin}=76.37\,\mu\mathrm{s}$, the single bin flux density noise scale is
\begin{equation}
\sigma_{S,{\rm bin,pol}}=\frac{\mathrm{SEFD}_{\rm pol}}{\sqrt{\Delta\nu_{\rm eff}\Delta t_{\rm bin}}},
\end{equation}
which gives $\sigma_{S,{\rm bin,RCP}}\simeq4.14$ Jy and $\sigma_{S,{\rm
bin,LCP}}\simeq3.85$ Jy.  A single pulse peak flux density is then estimated as
\begin{equation}
S_{{\rm pk,GP,pol}}\simeq (\mathrm{SNR})_{{\rm pk,pol}}\sigma_{S,{\rm bin,pol}}.
\end{equation}
For rup117 scan no0005, the RCP and LCP peak signal to noise ratios of 14.63 and 9.28
correspond to peak flux densities of 60.5 Jy and 35.7 Jy, respectively.  
Summing the two circular polarization streams gives an approximate
peak scale of $\simeq96$ Jy.  The maximum SNR event,
rup112 scan no0037, reaches $\simeq120$ Jy summed over the two circular
polarizations.

For a Gaussian bright pulse component, an approximate fluence follows from
\begin{equation}
\mathcal{F}_{{\rm GP,pol}}\simeq1.0645\,S_{{\rm pk,GP,pol}}W_{50},
\end{equation}
with $W_{50}$ in microseconds when the fluence is reported in Jy $\mu$s.  For rup117 scan
no0005, using the summed peak flux density $\simeq96$ Jy and $W_{50}\simeq0.61$ ms gives a
two polarization fluence of $\simeq6\times10^4$ Jy $\mu$s.  

The B1937+21 folded profile check in Table~\ref{tab:sefd} gives empirical 
single polarization SEFDs close to the adopted station values after the same retained bandwidth correction.

As a consistency check, the same scale can be compared with the mean
flux density. Inverting the strongest event by relative flux gives a scan local
clean mean profile peak of \(\simeq0.70\,{\rm Jy}\) in total intensity. With
\(S_{1400}=23\,{\rm mJy}\) and \(P\simeq0.1564\,{\rm s}\), the catalogue
flux density averaged over phase corresponds to a mean pulse fluence of
\(\simeq3.6\times10^{3}\,{\rm Jy}\,\mu{\rm s}\). The resulting equivalent
width is \(\simeq5.1\,{\rm ms}\), consistent with the broad mean profile with two components.

\begin{table}
\centering
\caption{SEFD consistency check from the PSR B1937+21 folded profiles. Total 5 hour observations, 10 scans.}
\label{tab:sefd}
\begin{tabular}{lrrrr}
\toprule
Profile & $N_{\rm scan}$ & SEFD, 128 MHz (Jy) & SEFD, eff. band (Jy) & Adopted (Jy) \\
\midrule
RCP & 10 & $390\pm33$ & $348\pm30$ & 365 \\
LCP & 10 & $379\pm40$ & $339\pm36$ & 340 \\
RCP+LCP & 10 & 272 & 243 & 249 \\
\bottomrule
\end{tabular}

\end{table}

\section{Discovery stage checks and auxiliary data}
\label{app:auxiliary}

\subsection{Trial DM and time resolution check}
\label{app:dm_check}

\begin{figure}
\centering
\includegraphics[width=\linewidth]{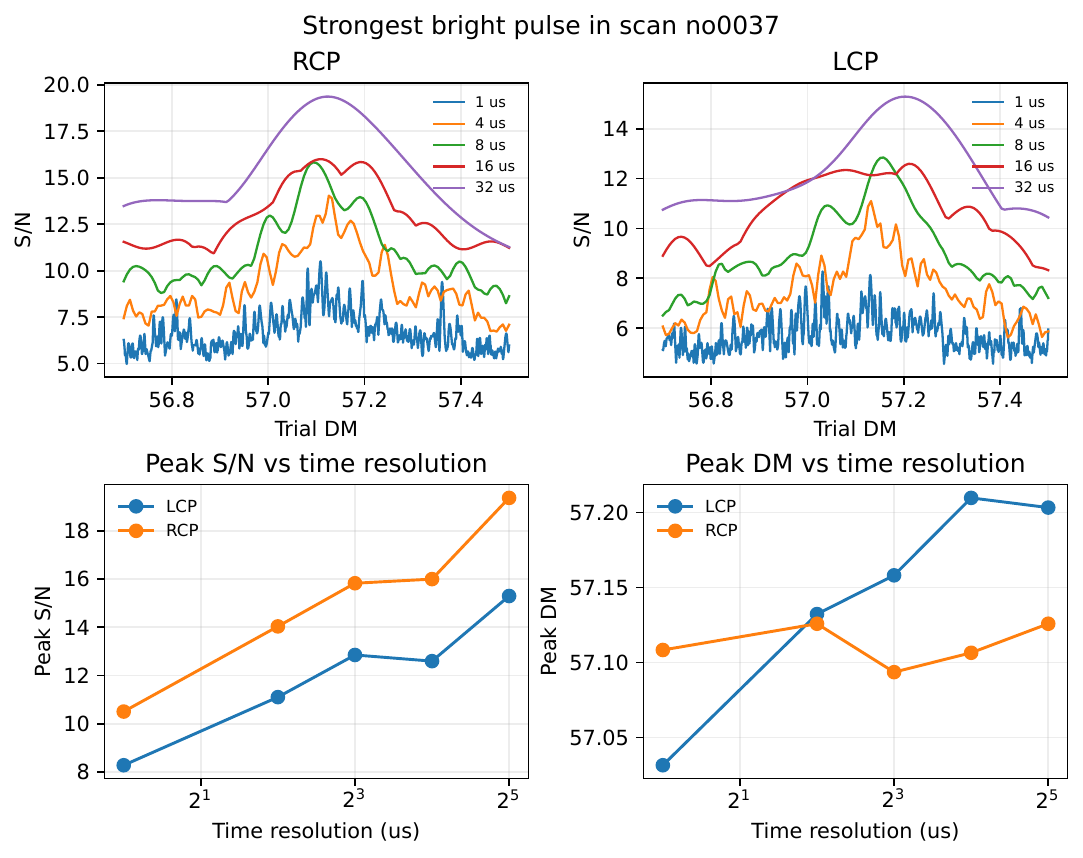}
\caption{Discovery stage coherent trial DM and time resolution check for the strongest bright pulse in scan no0037.  The event is recovered in both circular polarizations near the nominal DM of PSR B0355+54, and the maximum SNR increases toward coarser microsecond resolution.}
\label{edfig:dmcheck}
\end{figure}

The first bright pulse candidates were identified with the coherent single pulse search
pipeline previously developed for the Quasar VLBI pulsar programme
\cite{KurdubovMarshalov2022JAI}. 
The strongest event in scan no0037 was reprocessed over a grid of trial
dispersion measures at effective time resolutions of 1, 4, 8, 16 and
\(32\,\mu{\rm s}\) (Fig.~\ref{edfig:dmcheck}).
For each time resolution and circular
polarization, the pulse signal to noise ratio was measured as a function of coherent trial
DM.

The event is recovered in both circular polarizations near the expected dispersion measure
of PSR B0355+54.  The RCP curve gives a stable maximum near
$\mathrm{DM}\simeq57.09$--$57.13\,\mathrm{pc\,cm^{-3}}$. The LCP curve is noisier at
the finest resolution, but from 4 to 32 $\mu$s its maximum lies near
$\mathrm{DM}\simeq57.13$--$57.21\,\mathrm{pc\,cm^{-3}}$.  The peak signal to noise ratio
increases systematically toward coarser effective time resolution, from 10.5 to 19.4 in
RCP and from 8.3 to 15.3 in LCP.  
The strongest response does not occur at the finest tested
resolution, arguing against an instrumental artefact confined to one time bin
and motivating the folded archive analysis used for the population study.

\subsection{Extended figures and scan level summaries}
\label{app:extended_material}

\begin{figure}
\centering
\includegraphics[width=\linewidth]{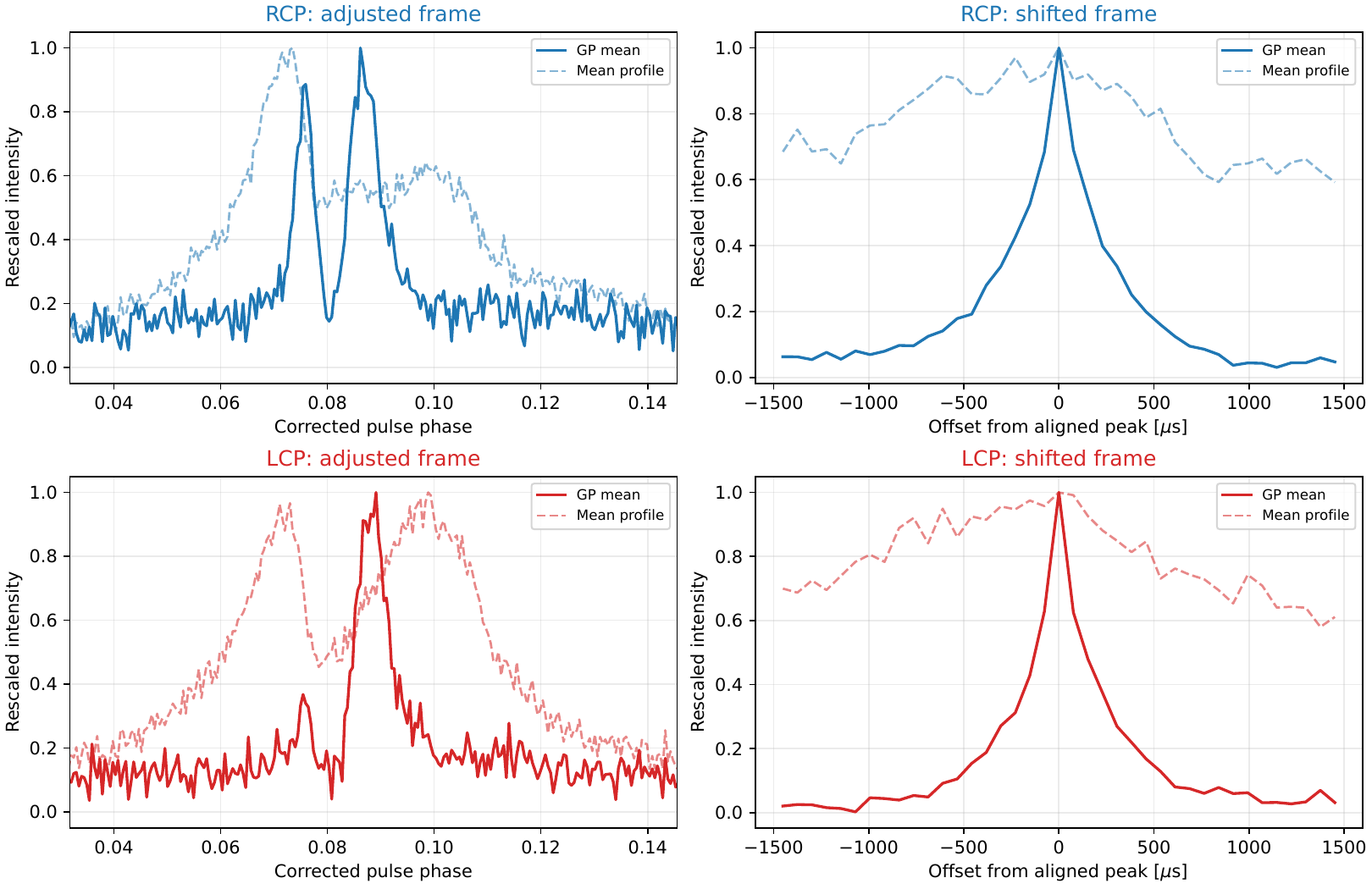}
\caption{Mean bright pulse profiles in the corrected and shifted frames.  Left panels show the clean mean profile and the mean of selected pulses in corrected pulse phase.  Right panels show the same comparison after shifting selected pulses to their own peaks.}
\label{edfig:gp_profiles}
\end{figure}

\begin{figure}
\centering
\includegraphics[width=\linewidth]{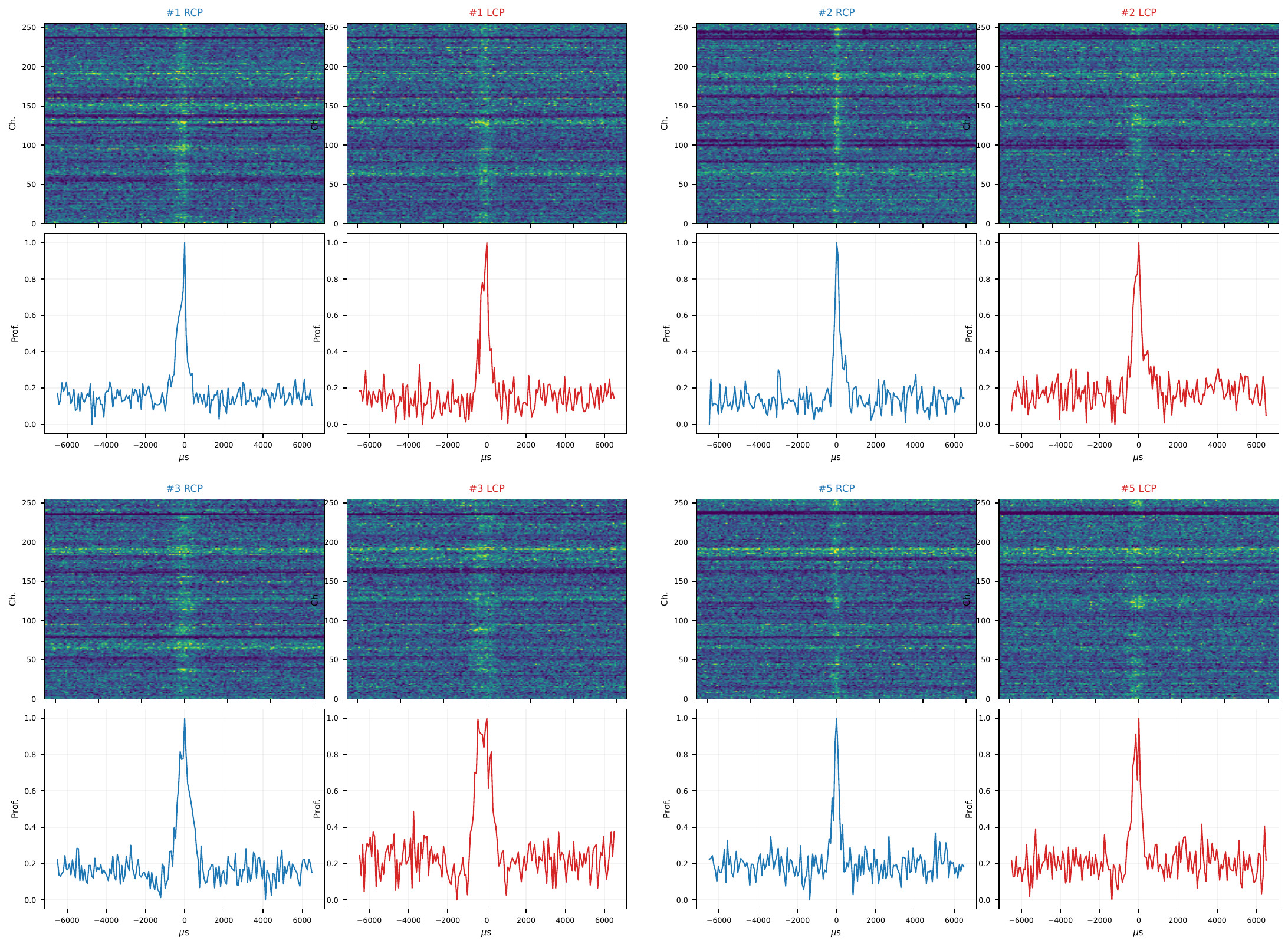}
\caption{Dynamic spectrum and profile cutouts for the strongest saved bright pulse events.  
The panels show RCP and LCP for each event before application of the fixed
channel mask, with the corresponding folded event profiles below.}

\label{edfig:event_gallery}
\end{figure}

\begin{table}
\centering
\caption{Parameters of the strongest saved bright pulse rotations. Each row gives the
RCP and LCP measurements for one GP.}
\label{edtab:strongest}
\begin{tabular}{llrrrrr}
\hline
Scan & Subint & RCP & LCP & RCP                                        & LCP                                        \\
     &        &  SNR   &  SNR    &     $S/\langle S\rangle$ &     $S/\langle S\rangle$ \\
\hline
rup112 no0037 & 10769 & 17.68 & 13.29 & 134.4 & 124.1 \\
rup117 no0001 & 7727 & 16.43 & 11.67 & 126.2 & 122.9 \\
rup117 no0005 & 118 & 14.63 & 9.28 & 149.7 & 121.7 \\
rup112 no0040 & 9670 & 12.84 & 9.78 & 108.3 & 104.7 \\
\hline
\end{tabular}
\end{table}

\clearpage



\begin{thebibliography}{43}
\expandafter\ifx\csname natexlab\endcsname\relax\def\natexlab#1{#1}\fi
\providecommand{\url}[1]{\texttt{#1}}
\providecommand{\href}[2]{#2}
\providecommand{\path}[1]{#1}
\providecommand{\DOIprefix}{doi:}
\providecommand{\ArXivprefix}{arXiv:}
\providecommand{\URLprefix}{URL: }
\providecommand{\Pubmedprefix}{pmid:}
\providecommand{\doi}[1]{\href{http://dx.doi.org/#1}{\path{#1}}}
\providecommand{\Pubmed}[1]{\href{pmid:#1}{\path{#1}}}
\providecommand{\bibinfo}[2]{#2}
\ifx\xfnm\relax \def\xfnm[#1]{\unskip,\space#1}\fi
\bibitem[{Bilous et~al.(2015)Bilous, Pennucci, Demorest and
  Ransom}]{Bilous2015B1821}
\bibinfo{author}{Bilous, A.V.}, \bibinfo{author}{Pennucci, T.T.},
  \bibinfo{author}{Demorest, P.}, \bibinfo{author}{Ransom, S.M.},
  \bibinfo{year}{2015}.
\newblock \bibinfo{title}{A broadband radio study of the average profile and
  giant pulses from {PSR B1821--24A}}.
\newblock \bibinfo{journal}{The Astrophysical Journal} \bibinfo{volume}{803},
  \bibinfo{pages}{83}.
\newblock \DOIprefix\doi{10.1088/0004-637X/803/2/83},
  \href{http://arxiv.org/abs/1412.7629}{\tt arXiv:1412.7629}.
\bibitem[{Cairns(2004)}]{Cairns2004GP}
\bibinfo{author}{Cairns, I.H.}, \bibinfo{year}{2004}.
\newblock \bibinfo{title}{Properties and interpretations of giant micropulses
  and giant pulses from pulsars}.
\newblock \bibinfo{journal}{The Astrophysical Journal} \bibinfo{volume}{610},
  \bibinfo{pages}{948--955}.
\newblock \DOIprefix\doi{10.1086/421756},
  \href{http://arxiv.org/abs/astro-ph/0404174}{\tt arXiv:astro-ph/0404174}.
\bibitem[{Cognard et~al.(1996)Cognard, Shrauner, Taylor and
  Thorsett}]{Cognard1996}
\bibinfo{author}{Cognard, I.}, \bibinfo{author}{Shrauner, J.A.},
  \bibinfo{author}{Taylor, J.H.}, \bibinfo{author}{Thorsett, S.E.},
  \bibinfo{year}{1996}.
\newblock \bibinfo{title}{Giant radio pulses from a millisecond pulsar}.
\newblock \bibinfo{journal}{The Astrophysical Journal Letters}
  \bibinfo{volume}{457}, \bibinfo{pages}{L81--L84}.
\bibitem[{Crawford et~al.(2013)Crawford, Altemose, Li and
  Lorimer}]{Crawford2013J0529}
\bibinfo{author}{Crawford, F.}, \bibinfo{author}{Altemose, D.},
  \bibinfo{author}{Li, H.}, \bibinfo{author}{Lorimer, D.R.},
  \bibinfo{year}{2013}.
\newblock \bibinfo{title}{Variability of the pulsed radio emission from the
  large magellanic cloud pulsar {PSR J0529--6652}}.
\newblock \bibinfo{journal}{The Astrophysical Journal} \bibinfo{volume}{762},
  \bibinfo{pages}{97}.
\newblock \DOIprefix\doi{10.1088/0004-637X/762/2/97},
  \href{http://arxiv.org/abs/1211.4531}{\tt arXiv:1211.4531}.
\bibitem[{Edwards et~al.(2006)Edwards, Hobbs and Manchester}]{EdwardsEtAl2006}
\bibinfo{author}{Edwards, R.T.}, \bibinfo{author}{Hobbs, G.B.},
  \bibinfo{author}{Manchester, R.N.}, \bibinfo{year}{2006}.
\newblock \bibinfo{title}{{TEMPO2}, a new pulsar timing package. ii. the timing
  model and precision estimates}.
\newblock \bibinfo{journal}{Monthly Notices of the Royal Astronomical Society}
  \bibinfo{volume}{372}, \bibinfo{pages}{1549--1574}.
\bibitem[{Ershov and Kuzmin(2003)}]{ErshovKuzmin2003B1112}
\bibinfo{author}{Ershov, A.A.}, \bibinfo{author}{Kuzmin, A.D.},
  \bibinfo{year}{2003}.
\newblock \bibinfo{title}{Detection of giant pulses from the pulsar {PSR
  B1112+50}}.
\newblock \bibinfo{journal}{Astronomy Letters} \bibinfo{volume}{29},
  \bibinfo{pages}{91--95}.
\bibitem[{Ershov and Kuzmin(2005)}]{ErshovKuzmin2005}
\bibinfo{author}{Ershov, A.A.}, \bibinfo{author}{Kuzmin, A.D.},
  \bibinfo{year}{2005}.
\newblock \bibinfo{title}{Detection of giant pulses in pulsar {PSR
  J1752+2359}}.
\newblock \bibinfo{journal}{Astronomy \& Astrophysics} \bibinfo{volume}{443},
  \bibinfo{pages}{593--597}.
\bibitem[{Hankins et~al.(2003)Hankins, Kern, Weatherall and
  Eilek}]{Hankins2003}
\bibinfo{author}{Hankins, T.H.}, \bibinfo{author}{Kern, J.S.},
  \bibinfo{author}{Weatherall, J.C.}, \bibinfo{author}{Eilek, J.A.},
  \bibinfo{year}{2003}.
\newblock \bibinfo{title}{Nanosecond radio bursts from strong plasma turbulence
  in the crab pulsar}.
\newblock \bibinfo{journal}{Nature} \bibinfo{volume}{422},
  \bibinfo{pages}{141--143}.
\bibitem[{Hobbs et~al.(2006)Hobbs, Edwards and Manchester}]{HobbsEtAl2006}
\bibinfo{author}{Hobbs, G.B.}, \bibinfo{author}{Edwards, R.T.},
  \bibinfo{author}{Manchester, R.N.}, \bibinfo{year}{2006}.
\newblock \bibinfo{title}{{TEMPO2}, a new pulsar-timing package. i. an
  overview}.
\newblock \bibinfo{journal}{Monthly Notices of the Royal Astronomical Society}
  \bibinfo{volume}{369}, \bibinfo{pages}{655--672}.
\bibitem[{Jiang et~al.(2024)}]{Jiang2024B0355Profile}
\bibinfo{author}{Jiang, S.}, et~al., \bibinfo{year}{2024}.
\newblock \bibinfo{title}{Profile variation in {PSR B0355+54} over a narrow
  frequency range}.
\newblock \bibinfo{journal}{Universe} \bibinfo{volume}{10},
  \bibinfo{pages}{416}.
\bibitem[{Johnston and Romani(2003)}]{JohnstonRomani2003}
\bibinfo{author}{Johnston, S.}, \bibinfo{author}{Romani, R.W.},
  \bibinfo{year}{2003}.
\newblock \bibinfo{title}{Giant pulses from {PSR B0540--69} in the large
  magellanic cloud}.
\newblock \bibinfo{journal}{The Astrophysical Journal Letters}
  \bibinfo{volume}{590}, \bibinfo{pages}{L95--L98}.
\bibitem[{Johnston et~al.(2004)Johnston, Romani, Marshall and
  Zhang}]{Johnston2004B0540}
\bibinfo{author}{Johnston, S.}, \bibinfo{author}{Romani, R.W.},
  \bibinfo{author}{Marshall, F.E.}, \bibinfo{author}{Zhang, W.},
  \bibinfo{year}{2004}.
\newblock \bibinfo{title}{Radio and x-ray observations of {PSR B0540--69}}.
\newblock \bibinfo{journal}{Monthly Notices of the Royal Astronomical Society}
  \bibinfo{volume}{355}, \bibinfo{pages}{31--36}.
\bibitem[{Kazantsev and Basalaeva(2022)}]{KazantsevBasalaeva2022}
\bibinfo{author}{Kazantsev, A.N.}, \bibinfo{author}{Basalaeva, M.Y.},
  \bibinfo{year}{2022}.
\newblock \bibinfo{title}{Low-frequency observations of giant pulses from
  ordinary pulsars}.
\newblock \bibinfo{journal}{Monthly Notices of the Royal Astronomical Society}
  \bibinfo{volume}{513}, \bibinfo{pages}{4332--4340}.
\bibitem[{Kinkhabwala and Thorsett(2000)}]{KinkhabwalaThorsett2000}
\bibinfo{author}{Kinkhabwala, A.}, \bibinfo{author}{Thorsett, S.E.},
  \bibinfo{year}{2000}.
\newblock \bibinfo{title}{Multifrequency observations of giant radio pulses
  from the millisecond pulsar {B1937+21}}.
\newblock \bibinfo{journal}{The Astrophysical Journal} \bibinfo{volume}{535},
  \bibinfo{pages}{365--372}.
\bibitem[{Knight(2007)}]{Knight2007J1823}
\bibinfo{author}{Knight, H.S.}, \bibinfo{year}{2007}.
\newblock \bibinfo{title}{A parkes radio telescope study of giant pulses from
  {PSR J1823--3021A}}.
\newblock \bibinfo{journal}{Monthly Notices of the Royal Astronomical Society}
  \bibinfo{volume}{378}, \bibinfo{pages}{723--729}.
\bibitem[{Knight et~al.(2005)Knight, Bailes, Manchester and
  Ord}]{Knight2005Search}
\bibinfo{author}{Knight, H.S.}, \bibinfo{author}{Bailes, M.},
  \bibinfo{author}{Manchester, R.N.}, \bibinfo{author}{Ord, S.M.},
  \bibinfo{year}{2005}.
\newblock \bibinfo{title}{A search for giant pulses from millisecond pulsars}.
\newblock \bibinfo{journal}{The Astrophysical Journal} \bibinfo{volume}{625},
  \bibinfo{pages}{951--956}.
\bibitem[{Knight et~al.(2006)Knight, Bailes, Manchester, Ord and
  Jacoby}]{Knight2006GBT}
\bibinfo{author}{Knight, H.S.}, \bibinfo{author}{Bailes, M.},
  \bibinfo{author}{Manchester, R.N.}, \bibinfo{author}{Ord, S.M.},
  \bibinfo{author}{Jacoby, B.A.}, \bibinfo{year}{2006}.
\newblock \bibinfo{title}{Green bank telescope studies of giant pulses from
  millisecond pulsars}.
\newblock \bibinfo{journal}{The Astrophysical Journal} \bibinfo{volume}{640},
  \bibinfo{pages}{941--949}.
\bibitem[{Kuiack et~al.(2020)}]{Kuiack2020B0950}
\bibinfo{author}{Kuiack, M.}, et~al., \bibinfo{year}{2020}.
\newblock \bibinfo{title}{Long-term study of extreme giant pulses from {PSR
  B0950+08} with aartfaac}.
\newblock \bibinfo{journal}{Monthly Notices of the Royal Astronomical Society}
  \bibinfo{volume}{497}, \bibinfo{pages}{846--854}.
\bibitem[{Kurdubov and Marshalov(2022)}]{KurdubovMarshalov2022JAI}
\bibinfo{author}{Kurdubov, S.L.}, \bibinfo{author}{Marshalov, D.A.},
  \bibinfo{year}{2022}.
\newblock \bibinfo{title}{Probing the new generation geodetic radio telescopes
  for {FRB} observations}.
\newblock \bibinfo{journal}{Journal of Astronomical Instrumentation}
  \bibinfo{volume}{11}, \bibinfo{pages}{2250015}.
\bibitem[{Kuzmin and Ershov(2004)}]{KuzminErshov2004}
\bibinfo{author}{Kuzmin, A.D.}, \bibinfo{author}{Ershov, A.A.},
  \bibinfo{year}{2004}.
\newblock \bibinfo{title}{Giant pulses in pulsar {PSR B0031--07}}.
\newblock \bibinfo{journal}{Astronomy \& Astrophysics} \bibinfo{volume}{427},
  \bibinfo{pages}{575--579}.
\bibitem[{Li et~al.(2016)}]{lwy+16}
\bibinfo{author}{Li, L.}, et~al., \bibinfo{year}{2016}.
\newblock \bibinfo{title}{Proper motions of 15 pulsars: A comparison between
  bayesian and frequentist algorithms}.
\newblock \bibinfo{journal}{Monthly Notices of the Royal Astronomical Society}
  \bibinfo{volume}{460}, \bibinfo{pages}{4011--4017}.
\bibitem[{Lundgren et~al.(1995)}]{Lundgren1995}
\bibinfo{author}{Lundgren, S.C.}, et~al., \bibinfo{year}{1995}.
\newblock \bibinfo{title}{Giant pulses from the crab pulsar: A joint radio and
  gamma-ray study}.
\newblock \bibinfo{journal}{The Astrophysical Journal} \bibinfo{volume}{453},
  \bibinfo{pages}{433--445}.
\bibitem[{Majid et~al.(2011)Majid, Naudet, Lowe and Kuiper}]{Majid2011}
\bibinfo{author}{Majid, W.A.}, \bibinfo{author}{Naudet, C.J.},
  \bibinfo{author}{Lowe, S.T.}, \bibinfo{author}{Kuiper, T.B.H.},
  \bibinfo{year}{2011}.
\newblock \bibinfo{title}{Statistical studies of giant pulse emission from the
  crab pulsar}.
\newblock \bibinfo{journal}{The Astrophysical Journal} \bibinfo{volume}{741},
  \bibinfo{pages}{53}.
\bibitem[{Manchester et~al.(2005)Manchester, Hobbs, Teoh and
  Hobbs}]{ManchesterEtAl2005}
\bibinfo{author}{Manchester, R.N.}, \bibinfo{author}{Hobbs, G.B.},
  \bibinfo{author}{Teoh, A.}, \bibinfo{author}{Hobbs, M.},
  \bibinfo{year}{2005}.
\newblock \bibinfo{title}{The australia telescope national facility pulsar
  catalogue}.
\newblock \bibinfo{journal}{The Astronomical Journal} \bibinfo{volume}{129},
  \bibinfo{pages}{1993--2006}.
\bibitem[{McGowan et~al.(2006)}]{McGowan2006B0355PWN}
\bibinfo{author}{McGowan, K.E.}, et~al., \bibinfo{year}{2006}.
\newblock \bibinfo{title}{Probing the pulsar wind nebula of {PSR B0355+54}}.
\newblock \bibinfo{journal}{The Astrophysical Journal} \bibinfo{volume}{647},
  \bibinfo{pages}{1300--1308}.
\bibitem[{McKee et~al.(2019)}]{McKee2019}
\bibinfo{author}{McKee, J.W.}, et~al., \bibinfo{year}{2019}.
\newblock \bibinfo{title}{A detailed study of giant pulses from {PSR B1937+21}
  using the large european array for pulsars}.
\newblock \bibinfo{journal}{Monthly Notices of the Royal Astronomical Society}
  \bibinfo{volume}{483}, \bibinfo{pages}{4784--4802}.
\bibitem[{Mitra et~al.(2008)Mitra, Sarala and Rankin}]{MitraSaralaRankin2008}
\bibinfo{author}{Mitra, D.}, \bibinfo{author}{Sarala, S.},
  \bibinfo{author}{Rankin, J.M.}, \bibinfo{year}{2008}.
\newblock \bibinfo{title}{Are partial cones aberrated cones?}, in:
  \bibinfo{booktitle}{40 Years of Pulsars: Millisecond Pulsars, Magnetars and
  More}, pp. \bibinfo{pages}{106--108}.
\bibitem[{Popov et~al.(2004)Popov, Soglasnov, Kondrat'ev and
  Kostyuk}]{PopovSoglasnov2004B1937Polarization}
\bibinfo{author}{Popov, M.V.}, \bibinfo{author}{Soglasnov, V.A.},
  \bibinfo{author}{Kondrat'ev, V.I.}, \bibinfo{author}{Kostyuk, S.V.},
  \bibinfo{year}{2004}.
\newblock \bibinfo{title}{Polarization observations of giant radio pulses from
  the millisecond pulsar {B1937+21} at a frequency of 600 mhz}.
\newblock \bibinfo{journal}{Astronomy Letters} \bibinfo{volume}{30},
  \bibinfo{pages}{95--99}.
\bibitem[{Popov and Stappers(2007)}]{PopovStappers2007}
\bibinfo{author}{Popov, M.V.}, \bibinfo{author}{Stappers, B.},
  \bibinfo{year}{2007}.
\newblock \bibinfo{title}{Statistical properties of giant pulses from the crab
  pulsar}.
\newblock \bibinfo{journal}{Astronomy \& Astrophysics} \bibinfo{volume}{470},
  \bibinfo{pages}{1003--1007}.
\bibitem[{Romani and Johnston(2001)}]{RomaniJohnston2001}
\bibinfo{author}{Romani, R.W.}, \bibinfo{author}{Johnston, S.},
  \bibinfo{year}{2001}.
\newblock \bibinfo{title}{Giant pulses from the millisecond pulsar
  {B1821--24}}.
\newblock \bibinfo{journal}{The Astrophysical Journal Letters}
  \bibinfo{volume}{557}, \bibinfo{pages}{L93--L96}.
\bibitem[{Shuygina et~al.(2019)}]{Shuygina2019}
\bibinfo{author}{Shuygina, N.}, et~al., \bibinfo{year}{2019}.
\newblock \bibinfo{title}{Russian vlbi network ``quasar'': Current status and
  outlook}.
\newblock \bibinfo{journal}{Geodesy and Geodynamics} \bibinfo{volume}{10},
  \bibinfo{pages}{150--156}.
\bibitem[{Soglasnov et~al.(2004)}]{Soglasnov2004}
\bibinfo{author}{Soglasnov, V.A.}, et~al., \bibinfo{year}{2004}.
\newblock \bibinfo{title}{Giant pulses from {PSR B1937+21} with widths less
  than 15 nanoseconds and brightness temperatures greater than $5\times10^{39}$
  k}.
\newblock \bibinfo{journal}{The Astrophysical Journal} \bibinfo{volume}{616},
  \bibinfo{pages}{439--451}.
\bibitem[{Staelin and Reifenstein(1968)}]{StaelinReifenstein1968}
\bibinfo{author}{Staelin, D.H.}, \bibinfo{author}{Reifenstein, E.C.},
  \bibinfo{year}{1968}.
\newblock \bibinfo{title}{Pulsating radio sources near the crab nebula}.
\newblock \bibinfo{journal}{Science} \bibinfo{volume}{162},
  \bibinfo{pages}{1481--1483}.
\bibitem[{Sun et~al.(2021)Sun, Yan and Wang}]{Sun2021J1047}
\bibinfo{author}{Sun, S.N.}, \bibinfo{author}{Yan, W.M.},
  \bibinfo{author}{Wang, N.}, \bibinfo{year}{2021}.
\newblock \bibinfo{title}{Detection of giant pulses in {PSR J1047--6709}}.
\newblock \bibinfo{journal}{Monthly Notices of the Royal Astronomical Society}
  \bibinfo{volume}{501}, \bibinfo{pages}{3900--3904}.
\bibitem[{Taylor et~al.(1993)Taylor, Manchester and
  Lyne}]{TaylorManchesterLyne1993}
\bibinfo{author}{Taylor, J.H.}, \bibinfo{author}{Manchester, R.N.},
  \bibinfo{author}{Lyne, A.G.}, \bibinfo{year}{1993}.
\newblock \bibinfo{title}{Catalog of 558 pulsars}.
\newblock \bibinfo{journal}{Astrophysical Journal Supplement Series}
  \bibinfo{volume}{88}, \bibinfo{pages}{529--568}.
\bibitem[{Tepedelenlioglu and Ogelman(2007)}]{Tepedelenlioglu2007B0355PWN}
\bibinfo{author}{Tepedelenlioglu, E.}, \bibinfo{author}{Ogelman, H.},
  \bibinfo{year}{2007}.
\newblock \bibinfo{title}{Discovery of extended emission around the pulsar
  {B0355+54}}.
\newblock \bibinfo{journal}{The Astrophysical Journal} \bibinfo{volume}{658},
  \bibinfo{pages}{1183--1187}.
\bibitem[{{van Straten} and Bailes(2011)}]{vanStratenBailes2011DSPSR}
\bibinfo{author}{{van Straten}, W.}, \bibinfo{author}{Bailes, M.},
  \bibinfo{year}{2011}.
\newblock \bibinfo{title}{{DSPSR}: Digital signal processing software for
  pulsar astronomy}.
\newblock \bibinfo{journal}{Publications of the Astronomical Society of
  Australia} \bibinfo{volume}{28}, \bibinfo{pages}{1--14}.
\bibitem[{{van Straten} et~al.(2012){van Straten}, Demorest and
  Oslowski}]{vanStraten2012PSRCHIVE}
\bibinfo{author}{{van Straten}, W.}, \bibinfo{author}{Demorest, P.},
  \bibinfo{author}{Oslowski, S.}, \bibinfo{year}{2012}.
\newblock \bibinfo{title}{Pulsar data analysis with {PSRCHIVE}}.
\newblock \bibinfo{journal}{Astronomical Research and Technology}
  \bibinfo{volume}{9}, \bibinfo{pages}{237--256}.
\bibitem[{{von Hoensbroech} et~al.(1998){von Hoensbroech}, Kijak and
  Krawczyk}]{vonHoensbroechKijakKrawczyk1998}
\bibinfo{author}{{von Hoensbroech}, A.}, \bibinfo{author}{Kijak, J.},
  \bibinfo{author}{Krawczyk, A.}, \bibinfo{year}{1998}.
\newblock \bibinfo{title}{On the high frequency polarization of pulsar radio
  emission}.
\newblock \bibinfo{journal}{Astronomy \& Astrophysics} \bibinfo{volume}{334},
  \bibinfo{pages}{571--584}.
\bibitem[{Weltevrede et~al.(2006a)Weltevrede, Stappers, Rankin and
  Wright}]{Weltevrede2006B0656RRAT}
\bibinfo{author}{Weltevrede, P.}, \bibinfo{author}{Stappers, B.W.},
  \bibinfo{author}{Rankin, J.M.}, \bibinfo{author}{Wright, G.A.E.},
  \bibinfo{year}{2006}a.
\newblock \bibinfo{title}{Is pulsar {B0656+14} a very nearby rotating radio
  transient?}
\newblock \bibinfo{journal}{The Astrophysical Journal Letters}
  \bibinfo{volume}{645}, \bibinfo{pages}{L149--L152}.
\newblock \DOIprefix\doi{10.1086/506346},
  \href{http://arxiv.org/abs/astro-ph/0606345}{\tt arXiv:astro-ph/0606345}.
\bibitem[{Weltevrede et~al.(2006b)Weltevrede, Wright, Stappers and
  Rankin}]{Weltevrede2006B0656}
\bibinfo{author}{Weltevrede, P.}, \bibinfo{author}{Wright, G.A.E.},
  \bibinfo{author}{Stappers, B.W.}, \bibinfo{author}{Rankin, J.M.},
  \bibinfo{year}{2006}b.
\newblock \bibinfo{title}{The bright spiky emission of pulsar {B0656+14}}.
\newblock \bibinfo{journal}{Astronomy \& Astrophysics} \bibinfo{volume}{458},
  \bibinfo{pages}{269--283}.
\newblock \DOIprefix\doi{10.1051/0004-6361:20065572},
  \href{http://arxiv.org/abs/astro-ph/0608023}{\tt arXiv:astro-ph/0608023}.
\bibitem[{Xu et~al.(2018)}]{Xu2018B0355Scint}
\bibinfo{author}{Xu, Y.H.}, et~al., \bibinfo{year}{2018}.
\newblock \bibinfo{title}{Interstellar scintillation observations for {PSR
  B0355+54}}.
\newblock \bibinfo{journal}{Monthly Notices of the Royal Astronomical Society}
  \bibinfo{volume}{476}, \bibinfo{pages}{5579--5590}.
\bibitem[{Zhuravlev et~al.(2013)}]{Zhuravlev2013B1937Polarization}
\bibinfo{author}{Zhuravlev, V.I.}, et~al., \bibinfo{year}{2013}.
\newblock \bibinfo{title}{Statistical and polarization properties of giant
  pulses of the millisecond pulsar {B1937+21}}.
\newblock \bibinfo{journal}{Monthly Notices of the Royal Astronomical Society}
  \bibinfo{volume}{430}, \bibinfo{pages}{2815--2821}.

\end{thebibliography}
\end{document}